\pdfoutput=1

\documentclass[11pt]{scrartcl}
\usepackage{times,amssymb,amsmath,graphicx}
\usepackage{sidecap,color}
\usepackage[bookmarks=false,pdffitwindow=false,pdfstartview={FitH},urlbordercolor={1 1 1}]{hyperref}

\usepackage{tgschola}
\usepackage[T1]{fontenc}
\usepackage[utf8]{inputenc}

\begin{document}
\title{\rmfamily
Nonstandard transition GUE-GOE for random matrices
and spectral statistics of~graphene nanoflakes \\}
\author{Adam Rycerz \\ 
  {\em Marian Smoluchowski Institute of Physics, }\\
  {\em Jagiellonian University, Łojasiewicza~11, } \\ 
  {\em 30--348 Kraków, Poland} }
\date{(Dated: June 20, 2016)}
\maketitle

\begin{abstract}
Spectral statistics of weakly-disordered triangular graphene flakes with zigzag edges are revisited. Earlier, we have found numerically that such systems may show spectral fluctuations of GUE, signalling the time-re\-ver\-sal symmetry breaking at zero magnetic field, accompanied by approximate twofold valley degeneracy of each energy level [Phys. Rev. B {\bf 85}, 245424 (2012)]. Atomic-scale disorder induces the scattering of charge carriers between the valleys and restores the spectral fluctuations of GOE. A simplified description of such a nonstandard GUE-GOE transition, employing the mixed ensemble of $4\times{}4$ real symmetric matrices was also proposed. Here we complement our previous study by analysing numerically the
spectral fluctuations of large matrices belonging the same mixed ensemble. Resulting scaling laws relate the ensemble parameter to physical size and the number of atomic-scale defects in graphene flake. A phase diagram, indicating the regions in which the signatures of GUE may by observable in the size-doping parameter plane, is presented. 
\end{abstract}

\section{\rmfamily Introduction}
The notion of emergent phenomena was coined out by P.W. Anderson in his milestone Science paper of 1979 \cite{And72}. In brief, emergence occurs when a complex system shows qualitatively different properties then its building blocks. Numerous examples of emergent systems studied in contemporary condensed matter physics, including high-temperature superconductors and heavy-fermion compounds \cite{Spa00}, are regarded as systems with spontaneous symmetry breaking \cite{And97}. A link between emergence and spontaneous symmetry breaking, however, does not seem to have a permanent character. In a wide class of electronic systems, such as semiconducting heterostructures containing a two-dimensional electron gas (2DEG), physical properties of itinerant electrons are substantially different than properties of free electrons (or electrons in atoms composing the system), and are also highly-tunable upon variation of external electromagnetic fields \cite{Ben91}. To give some illustration of this tunability, we only mention that electrons in GaAs heterostructures can be usually described by a standard Schr\"{o}dinger equation of quantum mechanics with the effective mass $m_{\rm eff}=0.067\,m_e$ (where $m_e$ is the free electron mass), whereas in extreme cases of quantum states formed in quantum Hall systems, effective quasiparticles may not even show the Fermi-Dirac statistics \cite{Ste10,Wil13}.

It is rather rarely noticed that graphene, a two-dimensional form of carbon just one atom tick \cite{Kat12}, also belongs to the second class of emergent systems (i.e., without an apparent spontaneous symmetry breaking) described briefly above. In a monolayer graphene, effective Hamiltonian for low-energy excitations has a Dirac-Weyl form, namely 
\begin{equation}
  \label{hameff}
  {\cal H}_{\rm eff}= 
  v_F\left[\,{\bf p}+e{\bf A}({\bf r},t)\,\right]
  \cdot{}\mbox{\boldmath$\sigma$}
  +U({\bf r},t), 
\end{equation}
where $v_F=10^6\,$m/s is the energy-independent Fermi velocity, $\mbox{\boldmath$\sigma$}=(\sigma_x,\sigma_y)$ with the Pauli matrices $\sigma_x$ and $\sigma_y$, ${\bf p}=-i\hbar(\partial_x,\partial_y)$ is the in-plane momentum operator, the electron charge is $-e$, and the external electromagnetic field is defined via scalar and vector potentials, $U({\bf r},t)$ and ${\bf A}({\bf r},t)$, with the in-plane position ${\bf r}=(x,y)$ and the time $t$.\footnote{Strictly speaking, ${\cal H}_{\rm eff}$ (\ref{hameff}) applies to quasiparticles near the $K$ valley in the dispersion relation. To obtain the effective Hamiltonian for other valley ($K'$) it is sufficient to substitute $\sigma_y\rightarrow{}-\sigma_y$.} In other words, the system build of nonrelativistic elements (carbon atoms at normal conditions) turns out to host ultrarelativistic quasiparticles, providing a beautiful example of an emergent phenomenon, which binds together two rather distant areas of relativistic quantum mechanics and condensed matter physics \cite{Abe10}. This observation applies generically to bilayer or multilayer graphenes \cite{Mac13}, as well as to HgTe/CdTe quantum wells \cite{Ber06}, although microscopic models describing such {\em other Dirac systems} are slightly different. It is also worth to mention so-called artificial graphenes, in which waves (of different kinds) obey their effective Dirac equations \cite{Sin11,Tor12,Gom12}.

A peculiar nature of Dirac fermions in graphene originates from the chiral structure of the Hamiltonian ${\cal H}_{\rm eff}$, accompanied by the fact that coupling to the external electromagnetic field is described by additive terms, which are linear in both scalar and vector potentials. A remarkable consequence of these facts is the quantization of the visible light absorption \cite{Nai08}, an unexpected macroscopic quantum effect recently found to have analogs in other Dirac systems \cite{Mac10,Shu13}, and even in a familiar graphite \cite{Sku10}. Another intriguing effect of this kind appears for dc conductivity of ballistic graphene \cite{Das11}. In the so-called pseudodiffusive transport regime, the conductance of a rectangular sample (with the width $W$ and the length $L$) scales as $G=\sigma_0\times{}W/L$ for $W\gg{}L$, where $\sigma_0=(4/\pi)e^2/h$ is the universal quantum value of the conductivity \cite{Two06,Dan08}, whereas the shot-noise power and all the other charge-transfer characteristics are indistinguishable from those of a classical diffusive conductor regardless the sample shape \cite{Ryc09}. At high magnetic fields, the pseudodiffusive charge transport is predicted theoretically to reappear for resonances with Landau levels in both monolayer \cite{Pra07} and bilayer graphene \cite{Rut14}. In the presence of disorder, a fundamental property of the Hamiltonian -- the time reversal symmetry (TRS) -- starts to play a decisive role. In particular, effective TRS in a single valley may be broken even in the absence of magnetic fields, leading to observable (and having the universal character) consequences for the conductance and spectral fluctuations \cite{Wur09,Ryc12}, as well as for the peculiar scaling behavior predicted for the conductivity \cite{Bar07,Ost10}.

Although the interest in graphene and other Dirac systems primarily focus on their potential applications \cite{Gei08,Gei09}, quite often linked to the nonstandard quantum description \cite{Abe10}, we believe that the fundamental perspective sketched in the above also deserves some attention. In the remaining part of this article, we first overview basic experimental, theoretical and numerical findings concerning signatures of quantum chaos in graphene and its nanostructures (Section~II). Next, we present our new numerical results concerning the additive random matrix model originally proposed in Ref.\ \cite{Ryc12} to describe a nonstandard GUE-GOE transition, accompanied by lifting out the valley degeneracy (Section~III). The consequences of these findings for prospective experiments on graphene nanoflakes, together with the phase diagram depicting the relevant matrix ensembles in the system size-doping plane, are described in Section~IV. The concluding remarks are given in Section~V.

\section{\rmfamily Gauge fields, fluctuations and chaos in nanoscale graphene structures}

Dirac fermions confined in graphene quantum dots \cite{Pon08} have provided yet another surprising situation, in which a piece of handbook knowledge needed a careful revision \cite{Lia14}.

Quantum chaotic behavior appears generically for systems, whose classical dynamics are chaotic, and manifests itself via the fact that energy levels show statistical fluctuations following those of Gaussian ensembles of random matrices \cite{Haa10}. In particular, if such a system posses the time-reversal symmetry (TRS), its spectral statistics follow the Gaussian orthogonal ensemble (GOE). A system with TRS and half-integer spin has the symplectic symmetry and, in turn, shows spectral fluctuations of the Gaussian symplectic ensemble (GSE). If TRS is broken, as in the presence of nontrivial gauge fields, and the system has no other antiunitary symmetry \cite{Rob86}, spectral statistics follow the Gaussian unitary ensemble (GUE). For a particular case of massless spin-1/2 particles, it was pointed out by Berry and Mondragon \cite{Ber87}, that the confinement may break TRS in a persistent manner (i.e., even in the absence of gauge fields), leading to the spectral fluctuations of GUE.

\begin{SCfigure}
  \centering
  \includegraphics[width=0.6\linewidth]{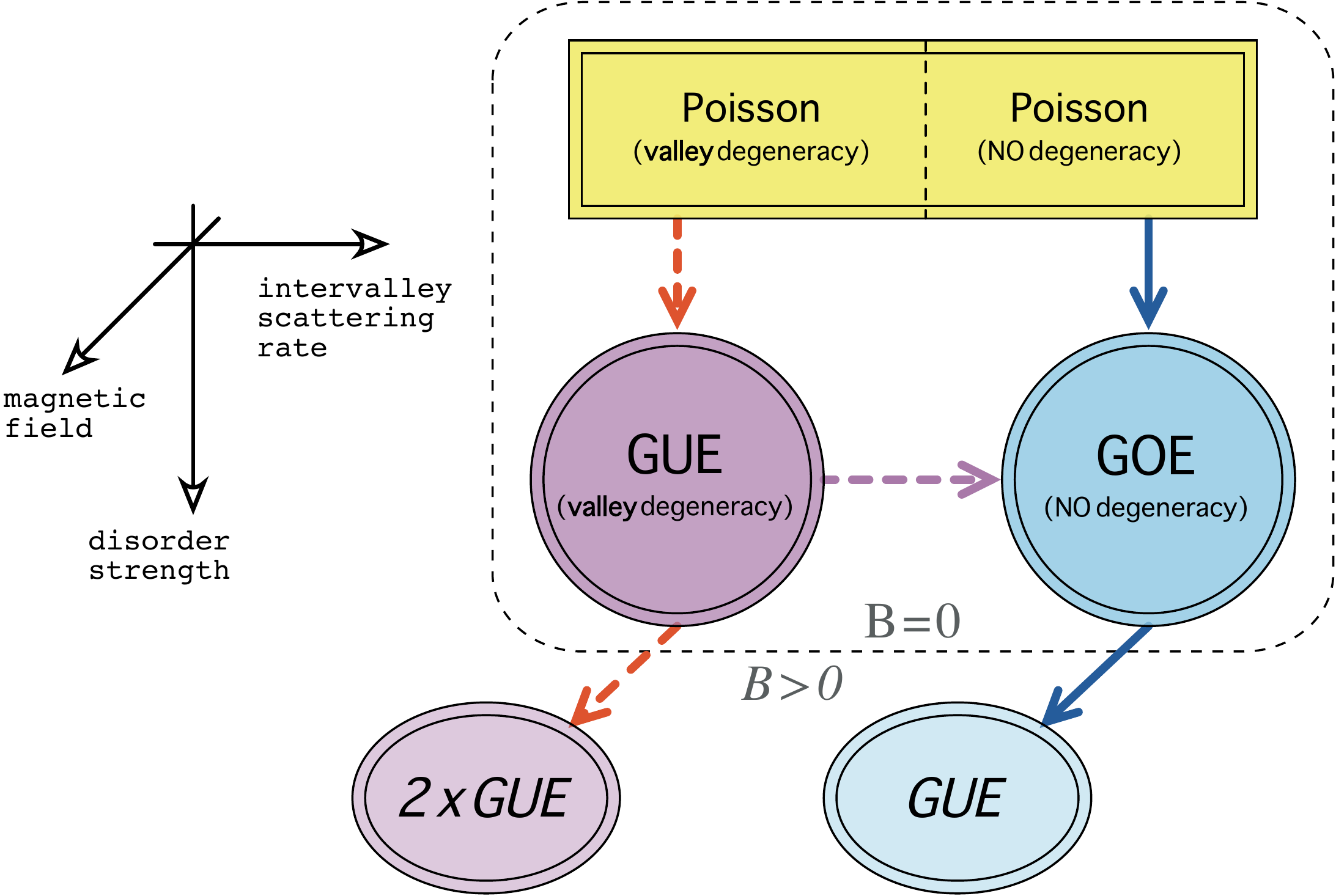}
  \caption{\label{gradyn}
    Transitions between symmetry classes and random matrix ensembles relevant for {\em closed} nanosystems in graphene characterized by the disorder strength, the intervalley scattering rate, and (optionally) placed in the weak magnetic field $B$. (Reprinted from Ref.\ \cite{Ryc12}.)
%Solid arrows in the right part indicate transitions reported in the literature prior to Ref.\ \cite{Ryc12}; dashed arrows indicate remaining transitions. 
  }
\end{SCfigure}

When applying the above symmetry classification to graphene nanosystems \cite{Wur09,Ryc12} one needs, however, to take into account that Dirac fermions in graphene appear in the two valleys, $K$ and $K'$, coupled by TRS. (In particular, real magnetic field break TRS and has the same sign in the two valleys, whereas the strain-induced gauge field preserves TRS and has opposite signs in the two valleys.) If the valley pseudospin is conserved, a special (symplectic) time-reversal symmetry (STRS) becomes relevant, playing a role of an effective TRS in a single valley \cite{Wur09}. Both real magnetic and strain-induced gauge fields may break STRS leading to the spectral fluctuations of GUE \cite{Ryc13}. As demonstrated numerically in Ref.\ \cite{Ryc12}, such fluctuations also appear for particular closed nanosystems in graphene in the presence of random scalar potentials slowly varying on the scale of atomic separation. Such nanosystems include equilateral triangles with zigzag or Klein edges, i.e., with terminal atoms belonging to one sublattice. Generic graphene nanoflakes with irregular edges show spectral fluctuations of GOE \cite{Wur09}, as strong intervalley scattering restores TRS in the absence of gauge fields (see Fig.\ \ref{gradyn}). In contrast, the boundary effects are suppressed in open graphene systems, for which signatures of the symplectic symmetry class were reported \cite{Pal12}.

It is worth mention here, that triangular graphene flakes, similar to studied theoretically in Ref.\ \cite{Ryc12}, have been recently fabricated \cite{Ham11,Oll12}. However, due to the hybridization with metallic substrates, quantum-dot energy levels in such systems are significantly broaden, making it rather difficult to determine the symmetry class via spectral statistics.

\section{\rmfamily Transition GUE-GOE for real symmetric matrices}
\subsection{\rmfamily Additive random-matrix models: Brief overview}
Additive random-matrix models are capable of reproducing the evolutions of spectral statistics in many cases when a~complex system undergoes transition to quantum chaos or transition between symmetry classes \cite{Haa10,Zyc93}. The discussion usually focus on the auxiliary random Hamiltonian of the form
\begin{equation}\label{admamod}
  H(\lambda)=\frac{H_0+\lambda{V}}{\sqrt{1+\lambda^2}}, 
\end{equation}
where $H_0=(H_0)^\dagger$  and $V=V^\dagger$ are members of different Gaussian ensembles\footnote{In order to describe transition to quantum chaos rather then transition between symmetry classes in a~chaotic system, one can choose $H_0$ to be a~diagonal random matrix, elements of which follow a~Gaussian distribution with zero mean and the variance $\langle(H_0)_{ij}^2\rangle=\delta_{ij}$.}, and the parameter $\lambda\in[0,\infty]$. 

For instance, if elements of $H_0$ are real numbers chosen to follow a~Gaussian distribution with zero mean and the variance $\langle(H_0)_{ij}^2\rangle=(1+\delta_{ij})/N$, where $\delta_{ij}$ is the Kronecker delta and $N$ is the matrix size, while elements of $V$ are complex numbers in which real and imaginary parts are generated independently according to Gaussian distribution with zero mean and the variance  $\langle(\mbox{Re}V_{ij})^2\rangle=(1+\delta_{ij})/2N$, $\langle(\mbox{Im}V_{ij})^2\rangle=(1-\delta_{ij})/2N$ (respectively), the Hamiltonian $H(\lambda)$ (\ref{admamod}) refers to transition GOE-GUE. For $N=2$, statistical distribution of the spacing between energy levels $S=|E_1-E_2|$ can be found analytically \cite{Len92}, and reads
\begin{equation}
  \label{lenzyc}
  P_{\rm GOE-GUE}(\lambda;S)=\sqrt{\frac{2+\lambda^2}{2}}Sc^2(\lambda) 
  \exp\left[-\frac{S^2c^2(\lambda)}{2}\right]\mbox{erf}
  \left[\frac{Sc(\lambda)}{\lambda}\right], 
\end{equation}
where $\mbox{erf}(x)$ is the error function, i.e., $\mbox{erf}(x)=(2/\sqrt{\pi})\int_0^x\exp(-t^2)dt$, and 
\begin{equation}
  \label{clamb}
  c(\lambda) = \sqrt{\frac{\pi(2\!+\!\lambda^2)}{4}}
  \left[1-\frac{2}{\pi}\left(\arctan\left(\frac{\lambda}{\sqrt{2}}\right)-\frac{\sqrt{2}\,\lambda}{2\!+\!\lambda^2}\right)\right]. 
\end{equation}
The above follows from the normalization condition
\begin{equation}
  \langle{}S\rangle=\int_0^{\infty}SP_{\rm GOE-GUE}(\lambda;S)dS=1
  \ \ \ \text{for}\ \ \ \ 0<\lambda<\infty. 
\end{equation}
The limiting forms of the spacing distribution given by Eqs.\ (\ref{lenzyc}) and (\ref{clamb}) are 
\begin{equation}
  \label{psgoe}
  P_{\rm GOE-GUE}(\lambda\rightarrow{}0;S)
  =\frac{\pi}{2}S\exp\left(-\frac{\pi{S^2}}{4}\right)
  \equiv{}P_{\rm GOE}(S),
\end{equation}
\begin{equation}
  \label{psgue}
  P_{\rm GOE-GUE}(\lambda\rightarrow{}\infty;S)
  =\frac{32}{\pi^2}S^2\exp\left(-\frac{4{S^2}}{\pi}\right)
  \equiv{}P_{\rm GUE}(S),
\end{equation}
coinciding with well-known Wigner surmises for GOE and GUE, respectively \cite{Haa10}. For $N\gg{}1$, it was also shown that actual spacing distributions obtained numerically can be approximated (with an astonishing accuracy) by $P_{\rm GOE-GUE}(\lambda_{\rm fit};S)$, where the empirical parameter $\lambda_{\rm fit}\propto{}\lambda\sqrt{N}$ \cite{Zyc93}. Similar scaling laws applies generically to all transitions between basic symmetry classes. 

Relatively recently, spectra of models employing self-dual random matrices have attracted some attention \cite{Sch12}. In such models, the matrix $H_0$ in Eq.\ (\ref{admamod}) is equivalent (up to a~unitary transformation) to the matrix having a~block structure 
\begin{equation}
\label{h0self}
  \tilde{H}_0=\left(\begin{array}{cc}
    C  & 0  \\
    0 & C^T  \\
  \end{array}\right), 
\end{equation}
where random matrix $C$ is an $N\times{}N$ member of one of Gaussian ensembles, $C^T$ denotes the transposition of $C$. The matrix $V$ in Eq.\ (\ref{admamod}) is a~generic $2N\times{}2N$ member of the other ensemble (hereinafter, we redefine the $H(\lambda)$ size as $2N$). In turn, for $\lambda=0$, each eigenvalue is doubly degenerate. For $\lambda\neq{}0$, we have the degeneracy splitting accompanied by transition between selected symmetry classes. Even in the simplest case of $N=2$, closed-form analytic expressions for level-spacing distributions corresponding to arbitrary $0<\lambda<\infty$ are missing. 
The approach presented in Ref.\ \cite{Sch12} employs the relevant expressions for joint probability densities for eigenvalues \cite{Meh04}, allowing one to express level-spacings distribution in terms of two-dimensional integrals to be evaluated numerically.

In the remaining part of Section, we focus on the transition between self-dual GUE to GOE, show that the corresponding Hamiltonian $H(\lambda)$, and can be represented as real-symmetric random matrix, and present our empirical expressions approximating spacing distributions obtained numerically. 

%\noindent
%[{\color{blue}A MOZE COS O WEKTORACH WLASNYCH?}]

\subsection{\rmfamily Self-dual GUE to GOE via \boldmath{$4\!\times{}\!4$} real-symmetric matrices}
We focus now on the situation, when the matrix $C$ in Eq.\ (\ref{h0self}) is chosen to be an $N\times{}N$  member of GUE, whereas $V$ in Eq.\ (\ref{admamod}) is a~$2N\times{}2N$ member of GOE. 

For $N=2$, the matrix $\tilde{H}_0$ can be written as
\begin{equation}
\label{h0sd4x4}
  \tilde{H}_0^{4\times{}4}=\left(\begin{array}{cccc}
    a    & c+id   & 0    & 0    \\
    c-id & b      & 0    & 0    \\
    0    & 0      & a    & c-id \\
    0    & 0      & c+id & b    \\
  \end{array}\right), 
\end{equation}
where $a$ and $b$ are real random numbers following Gaussian distribution with zero mean and the variance $\langle{}a^2\rangle=\langle{}b^2\rangle=1/2$, whereas $c$ and $d$ are real random numbers following Gaussian distribution with zero mean and the variance $\langle{}c^2\rangle=\langle{}d^2\rangle=1/4$. Exchanging the second row with the third row, as well as the second column with the third column, we find the matrix $\tilde{H}_0^{4\times{}4}$ is equivalent, up to an orthogonal transformation, to
\begin{equation}
\label{h0sd4x4step1}
  \tilde{H}_0^{4\times{}4}\overset{O}{\longleftrightarrow}\left(
    \begin{array}{cccc}
      a    & 0   & c+id    & 0    \\
      0 & a      & 0    & c-id    \\
      c-id    & 0      & b    & 0 \\
      0    & c+id      & 0 & b    \\      
    \end{array}\right). 
\end{equation}
The matrix on the right-hand side of Eq.\ (\ref{h0sd4x4step1}) is self-dual, and can be further transformed as
\begin{equation}
  \label{uh0ud}
  U\left(
    \begin{array}{cccc}
      a    & 0   & c+id    & 0    \\
      0 & a      & 0    & c-id    \\
      c-id    & 0      & b    & 0 \\
      0    & c+id      & 0 & b    \\      
    \end{array}\right)U^\dagger
  = \left(
    \begin{array}{cccc}
      a    & 0   & c    & d    \\
      0 & a      & -d    & c    \\
      c    & -d      & b    & 0 \\
      d    & c      & 0 & b    \\      
    \end{array}\right), 
\end{equation}
where
\begin{equation}
  U=\frac{1}{\sqrt{2}} {1\!\!1}_{2\times{}2}\otimes\left(
    \begin{array}{cc} 
      1 & 1 \\
      i & -i \\
    \end{array}\right) 
  = \frac{1}{\sqrt{2}} \left(
    \begin{array}{cccc}
      1    & 1   & 0    & 0    \\
      i & -i      & 0    & 0    \\
      0    & 0      & 1    & 1 \\
      0    & 0      & i & -i    \\      
    \end{array}\right). 
\end{equation}
Exchanging the second with the third row and column in the rightmost matrix in Eq.\ (\ref{uh0ud}) we arrive to
\begin{equation}
  \label{h0block4x4}
  H_0^{4\times{}4} = \left(
    \begin{array}{cccc}
      a & c  & 0  & d  \\
      c & b  & -d & 0  \\
      0 & -d & a  & c  \\
      d & 0  & c  & b  \\      
    \end{array}\right)
  = \left(
    \begin{array}{cc} 
      A  & B \\
      -B & A \\
    \end{array}\right),    
\end{equation}
where the blocks $A$ and $B$ are real-symmetric ($A^T=A$) and skew-symmetric ($B^T=-B$) random matrices. 

Spectral statistics of the Hamiltonian $H(\lambda)=\left(H_0^{4\times{}4}+\lambda{V^{4\times{}4}}\right)/\sqrt{1+\lambda^2}$, with $V^{4\times{}4}$ being a~$4\times{}4$ GOE matrix, were thoroughly studied before \cite{Ryc12appd}. Here we revisit our findings, before discussing spectra of larger matrices in next subsection. 

The nearest-neighbor spacings distribution can be approximated by
\begin{equation}
  \label{psakap}
  P({\alpha,\kappa};S) = \frac{\alpha P_{\rm GOE}(\alpha{S})
    +\beta(\alpha) P_{\rm GOE-GUE}(\kappa;\beta(\alpha) S)}{2},
\end{equation}
with $\beta(\alpha)=\alpha/(2\alpha-1)$, $P_{\rm GOE}({S})$ given by Eq.\ (\ref{psgoe}), $P_{\rm GOE-GUE}(\kappa;{S})$ given by Eq.\ (\ref{lenzyc}), and the parameters $\alpha$ and $\kappa$ which can be approximated by empirical functions
\begin{equation}
  \label{alp4x4}
  \alpha\approx{}
\overline{\alpha}_{4\times{}4}(\lambda)=1.118\times\left[\sqrt[3]{1+(0.60/\lambda)^3}\right]^{0.98},
\end{equation}
and 
\begin{equation}
  \label{kap4x4}
  \kappa\approx{}
  \overline{\kappa}_{4\times{}4}(\lambda)=\sqrt{\left(\frac{1+\lambda^{-2}}{1+(0.33)^{-2}}\right)^{0.29}-1}.
\end{equation}
Eqs.\ (\ref{alp4x4}) and (\ref{kap4x4}) represents simplified versions of the corresponding formulas given in Ref.\ \cite{Ryc12appd}. A~comparison with the numerical will be given later in this section.

\subsection{\rmfamily Self-dual GUE to GOE via \boldmath{$2N\!\times{}\!2N$} real-symmetric matrices}
We consider now the case of large random matrices ($N\gg{}1$). A~generalization of the reasoning presented in previous subsection brought as to the unperturbed Hamiltonian $H_0$ with the block structure as given by the last equality in Eq.\ (\ref{h0block4x4}), but  $A=A^T$ and $B=-B^T$ are now $N\times{}N$ random matrices. The elements of each block are independently generated according to a Gaussian distribution with zero mean and the variance $\mbox{Var}(A_{ij})=(1+\delta_{ij})/2N$ and $\mbox{Var}(B_{ij})=(1-\delta_{ij})/2N$, respectively. In turn, $H_0$ can be unitary mapped onto the matrix $\tilde{H}_0$ given by Eq.\ (\ref{h0self}) with $C$ being an $N\times{}N$ member of GUE. The additive random-matrix model $H(\lambda)$ is complemented with the perturbation $V$ being an $2N\times{}2N$ member of GOE. 

\begin{figure}[!p]
\centerline{ \includegraphics[width=0.8\linewidth]{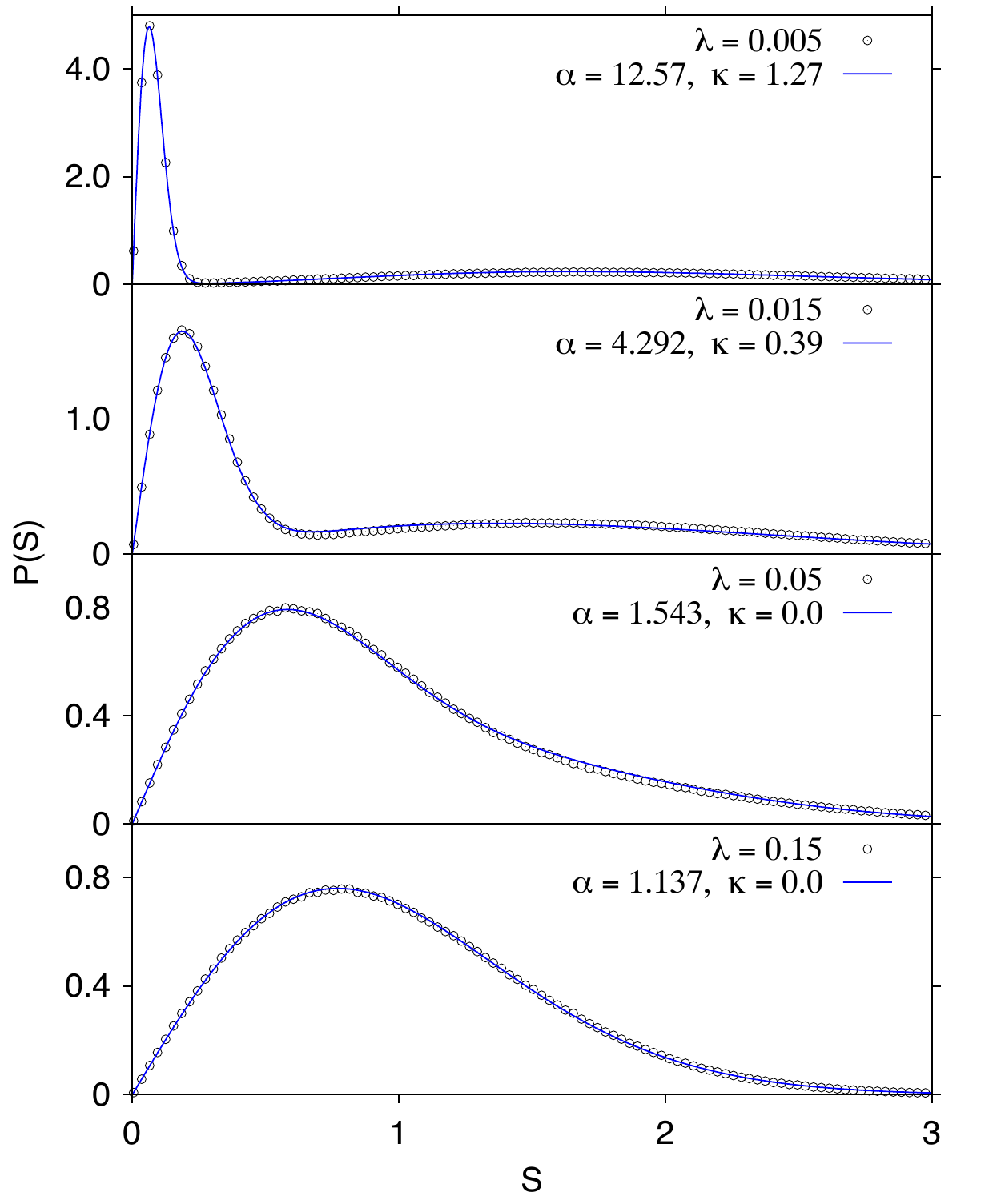} }
\caption{ \label{psfit400}
  Level-spacing distributions for $10^5$ randomly-generated Hamiltonians $H(\lambda)$ with the size $2N=400$ (datapoints). The scaling parameters $\lambda$ is varied between the panels. The least-squares fitted functions $P(\alpha,\kappa;S)$ defined by Eq.\ (\ref{psakap}) are also shown (solid lines). 
}
\end{figure}

Ensembles of large pseudo-random Hamiltonians $H(\lambda)$ were generated and diagonalized numerically, to check whether the standard scaling law $\lambda_{\rm fit}\simeq{}(2N)^{1/2}\lambda$ \cite{Len91} applies to spacings distribution of such matrices. Our presentation is limited to the matrix sizes $2N=200$, $400$, and $1000$; the statistical ensemble consists of the total amount of $10^6$, $10^5$, or $10^4$ matrices (respectively), same for each considered value of the parameter $\lambda$. To avoid the boundary effects, we limit our numerical study to about $30\%$ of the energy levels such that $|E|\leqslant{}0.5$. Selected examples are presented in Fig.\ \ref{psfit400}. 

\begin{figure}[!p]
\centerline{\includegraphics[width=\linewidth]{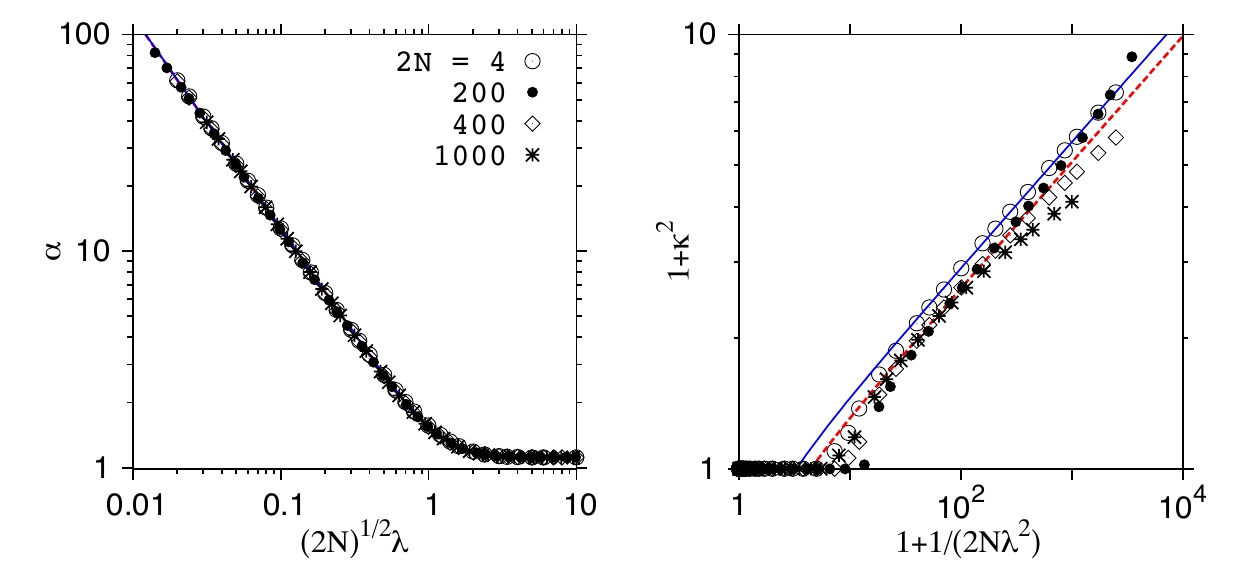}}
\caption{\label{fitparlarge}
  Least-squares fitted parameters of $P(\alpha,\kappa;S)$ (\ref{psakap}) for different values of $2N$ as functions of the scaled model parameter $(2N)^{1/2}\lambda$ (datapoints). The empirical relations $\overline{\alpha}_{4\times{}4}(\lambda)$ (\ref{alp4x4}) and  $\overline{\kappa}_{4\times{}4}(\lambda)$ (\ref{kap4x4}) valid for $2N=4$ are shown with blue solid lines; the relations $\overline{\alpha}_{N\gg{}1}(\lambda)$ (\ref{alplarg}) and  $\overline{\kappa}_{N\gg{}1}(\lambda)$ (\ref{kaplarg}) for large matrices are shown with red dashed lines.
}
\end{figure}

\begin{figure}[!p]
\centerline{\includegraphics[width=0.8\linewidth]{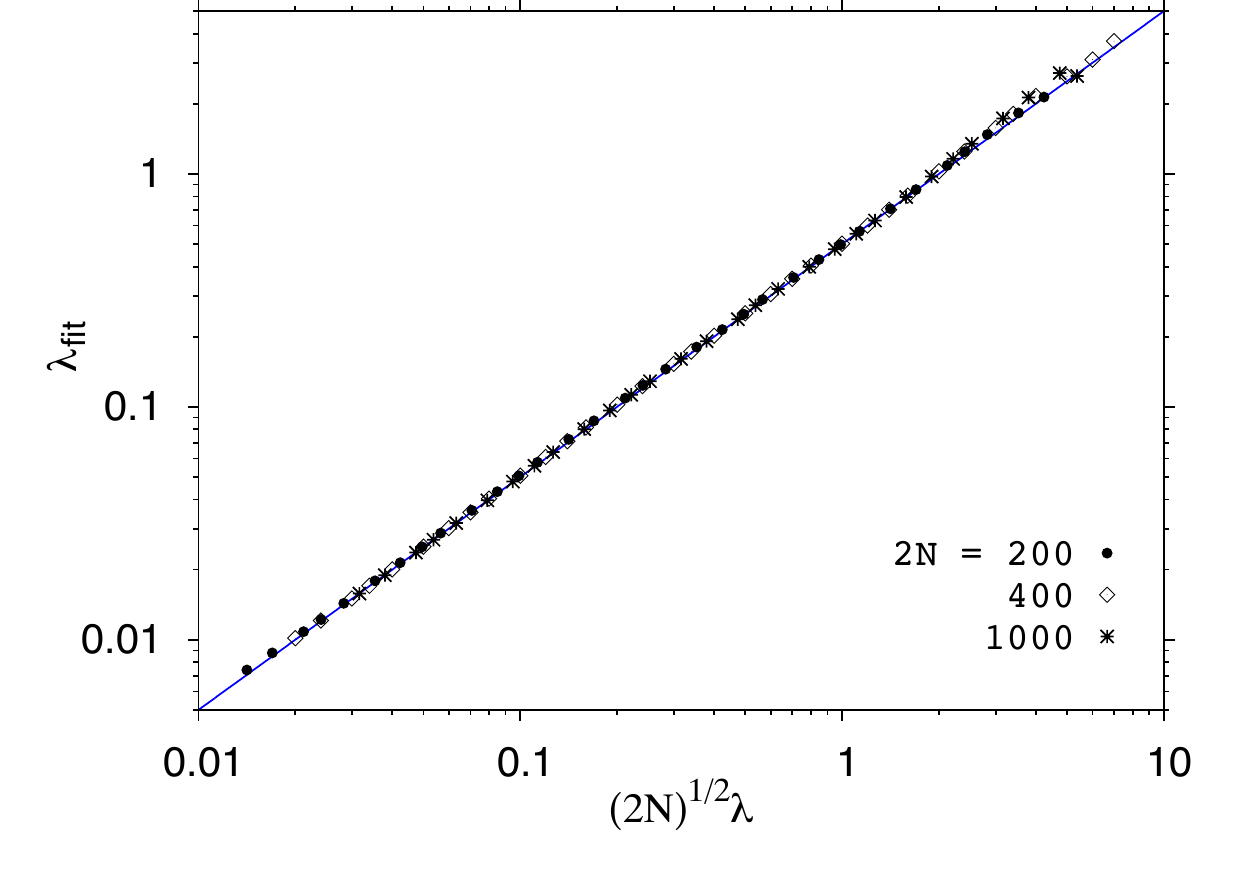}}
\caption{\label{lamfitlarge}
  Scaling law for the best fitted parameters $\tilde{\lambda}=\lambda_{\rm fit}$ in the distribution $P_{\overline{\alpha},\overline{\kappa}}(\tilde{\lambda},S)$ (\ref{psoveralka}) approximating $P(S)$ obtained numerically for random Hamiltonians $H(\lambda)$ with $2N=200$, $400$ and $1000$. [See the main text for details.] Blue solid line marks $\lambda_{\rm fit}=\lambda$. 
}
\end{figure}

We find that nearest-neighbor level spacings of large matrix $H(\lambda)$ follow the empirical distribution having the general form as given by Eq.\ (\ref{psakap}) 
\begin{equation}
  \label{psoveralka}
  P_{\overline{\alpha},\overline{\kappa}}(\tilde{\lambda},S) = 
  P(\overline{\alpha}_{N\gg{}1}(\tilde{\lambda}),\overline{\kappa}_{N\gg{}1}(\tilde{\lambda});S),
\end{equation}
with the empirical relations of Eqs.\ (\ref{alp4x4}) and (\ref{kap4x4}) [see blue solid lines in Fig.\ \ref{fitparlarge}] now replaced by
\begin{equation}
  \label{alplarg}
  \overline{\alpha}_{N\gg{}1}(\tilde{\lambda})= 
  1.114\times\left[\sqrt[3]{1+(0.60/\tilde{\lambda})^3}\right]^{0.98},
\end{equation}
and
\begin{equation}
  \label{kaplarg}
  \overline{\kappa}_{N\gg{}1}(\tilde{\lambda})= 
  \begin{cases}
  \sqrt{\left[\frac{ 1+\tilde{\lambda}^{-2}}{ 1+(0.27)^{-2}}\right]^{0.29}-1} & \quad \text{if}\ \ \tilde{\lambda}<0.27, \\
  0 & \quad \text{if}\ \ \tilde{\lambda}\geqslant{}0.27. \\
  \end{cases}
\end{equation}
The above formula are mark in Fig.\ \ref{fitparlarge} with red dashed lines. We also find that the scaling law $\tilde{\lambda}=\lambda_{\rm fit}\simeq{}(2N)^{1/2}\lambda$ [with $\lambda$ being the original parameter of $H(\lambda)$] is satisfied for the matrices considered with a~surprising accuracy (see Fig.\ \ref{lamfitlarge}).

%\noindent
%[{\color{blue}MOŻE COŚ O $\Delta_3(L)?$}]

\section{\rmfamily Consequences for graphene nanoflakes}

\subsection{\rmfamily Level-spacing distributions revisited}
In this subsection, the empirical distribution $P_{\overline{\alpha},\overline{\kappa}}(\lambda,S)$ (\ref{psoveralka})
with least-square fitted $\lambda=\lambda_{\rm fit}$ is utilized to rationalize level-spacing distributions for triangular graphene nanoflakes with zigzag edges. 

At zero magnetic field, the tight-binding Hamiltonian for weakly-disordered gra\-phe\-ne can be written as
\begin{equation}
  \label{hamtba}
  {\cal H}_{\rm TBA}= 
  \sum_{\langle{ij}\rangle}\left[\,t_{ij}|{i}\rangle\langle{j}| +
    {\rm h.c.}\,\right] \\
  + \sum_i\left[\,M_V({\bf r}_i) + 
    U_{\rm imp}({\bf r}_i)\,\right]|{i}\rangle\langle{i}|,
\end{equation}
where $t_{ij}=-t$ if the orbitals $|i\rangle$ and $|j\rangle$ are nearest neighbors on the honeycomb lattice (with  $t=\frac{2}{3}\sqrt{3}\hbar{v}_F/a\approx{}3\,$eV, and $a=0.246\,$nm being the lattice spacing), otherwise $t_{ij}=0$. (The symbol $\sum_{\langle{ij}\rangle}$ denotes that each pair $\langle{ij}\rangle$ is counted only once.) The terms $M_V({\bf r}_i)$ and $U_{\rm imp}({\bf r}_i)$ represent the potentials abruptly and slowly varying on the scale of atomic separation (respectively). Here we put $M_V({\bf r}_i)=0.7\,t$ if ${\bf r}_i$ is the outermost atom position at zigzag edge, otherwise $M_V({\bf r}_i)=0$. The random contribution $U_{\rm imp}({\bf r}_i)$ is generated in as follows: First, we randomly choose $N_{\rm imp}$ lattice sites ${\bf R}_n$ ($n=1,\dots,N_{\rm imp}$) out of $N_{\rm tot}$. Next, the amplitudes $U_n\in(-\delta,\delta)$ are randomly generated. Finally, the potential is smoothed over a distance $\xi=\sqrt{3}\,a$ by convolution with a Gaussian, namely
\begin{equation} \label{uimper}
  U_{\rm imp}({\bf r})=
  \sum_{n=1}^{N_{\rm imp}}U_n\exp\left(-\frac{|{\bf r}-{\bf R}_n|^2}{2\xi^2}\right).
\end{equation}

A~model of substrate-induced disorder, constituted by Eqs.\ (\ref{hamtba}) and (\ref{uimper}), was widely used to reproduce numerically several transport properties of disordered graphene samples \cite{Ryc07,Lew08,Hor09,Wur11b}. Here we revisit the spectra of closed graphene flakes considered in Ref.\ \cite{Ryc12}, within a~simplified empirical model $P_{\overline{\alpha},\overline{\kappa}}(\lambda_{\rm fit},S)$ (\ref{psoveralka}), in order discuss the consequences for prospective experimental observation of the zero-field time-reversal symmetry breaking in such systems. 

A~compact measure of the disorder strength is given by the dimensionless correlator
\begin{equation} 
  \label{knoddef}
  K_0=\frac{\cal A}{\left(\hbar{v}_F\right)^2}\frac{1}{N_{\rm tot}^2}
  \sum_{i=1}^{N_{\rm tot}}\sum_{j=1}^{N_{\rm tot}}\left\langle
    {U_{\rm imp}({\bf r}_i)U_{\rm imp}({\bf r}_j)}\right\rangle,
\end{equation}
where the system area ${\cal A}=\frac{1}{4}\sqrt{3}N_{\rm tot}a^2$, and the averaging takes place over possible realizations of the disorder in Eq.\ (\ref{uimper}). For $\xi\gg{}a$ Eq.\ (\ref{knoddef}) leads to
\begin{equation} 
  \label{knodval}
  K_0 = \frac{64\pi^2\sqrt{3}}{27}\,\frac{N_{\rm imp}}{N_{\rm tot}}
  \left(\frac{\delta}{t}\right)^2\left(\frac{\xi}{a}\right)^4.  
\end{equation}
For $\xi=\sqrt{3}\,a$, used for numerical demonstration in the remaining of this article, Eq.\ (\ref{knodval}) still provides a~good approximation of the actual value of $K_0$ and can be rewritten as
\begin{equation} 
  \label{knodxi3}
  K_0\approx{}364.7\times
  \frac{N_{\rm imp}}{N_{\rm tot}}\left(\frac{\delta}{t}\right)^2.
\end{equation}

\begin{figure}[!t]
\centerline{\includegraphics[width=\linewidth]{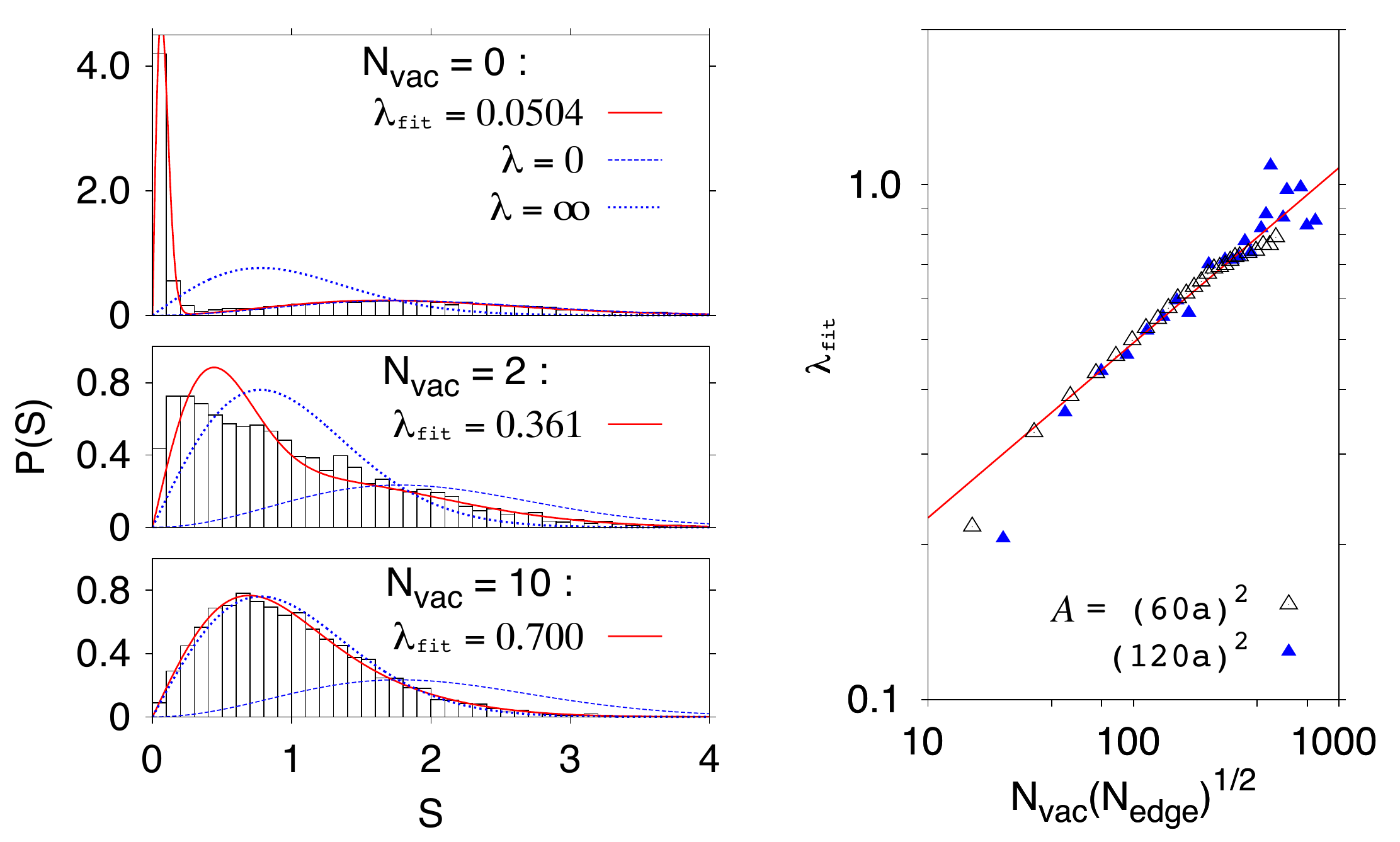}}
\caption{\label{psnvactri}
  {\em Left:} Level-spacing distributions $P(S)$ for triangular graphene nanoflakes with zigzag edges. The flake area is ${\cal A}\approx{}(120\,a)^2$, the disorder strength is $K_0\approx{}0.125$, the number of edge vacancies $N_{\rm vac}$ is varied between the panels. Numerical results (replotted from Ref.\ \cite{Ryc12}) are
shown with black solid lines. The other lines correspond to empirical
distributions  $P_{\overline{\alpha},\overline{\kappa}}(\lambda,S)$ (\ref{psoveralka}) with $\lambda=\lambda_{\rm fit}$ (red solid line), $\lambda=0$ (blue dashed line) or $\lambda=\infty$ (blue dotted line). 
  {\em Right:} Least-squares fitted parameters for different numbers of edge vacancies $1\leqslant{}N_{\rm vac}\leqslant{}30$ and the flake areas  ${\cal A}\approx{}(60\,a)^2$ (open symbols) and  ${\cal A}\approx{}(120\,a)^2$ (closed symbols), corresponding to the total number of terminal atoms $N_{\rm edge}=270$ and $540$ (respectively). Solid line depicts the approximating power-law relation given by Eq.\ (\ref{lamfitnvac}). 
}
\end{figure}

The numerical results are presented in Fig.\ \ref{psnvactri}, where we have fixed the remaining disorder parameters at $\delta/t=0.1$ and $N_{\rm imp}/N_{\rm tot}=0.034$ leading to $K_0=0.125$.\footnote{The disorder parameters are actually same as in Figs.\ 8 and 9 of Ref.\ \cite{Ryc12}, where we have mistakenly omitted the factor $\pi$ in the numerical evaluation of $K_0$.} Level-spacing distributions $P(S)$ obtained numerically for triangular nanoflakes with zigzag edges [see left panels in Fig.\ \ref{psnvactri}, black solid lines] are replotted from Ref.\ \cite{Ryc12}, where we used approximately $1500$ energy levels with energies $0.1\leqslant{}|E|/t\leqslant{}0.5$ out of the total number of $N_{\rm tot}(N_{\rm vac})=32758-N_{\rm vac}$ (corresponding the flake area ${\cal A}=(120\,a)^2$), with $N_{\rm vac}$ being the number of vacancies, randomly distributed along the system boundary. Typically, best-fitted parameters $\lambda=\lambda_{\rm fit}$ of the simplified distribution $P_{\overline{\alpha},\overline{\kappa}}(\lambda,S)$ (\ref{psoveralka}) coincides with given in Ref.\ \cite{Ryc12} up to a~second decimal place. New values of $\lambda_{\rm fit}$ for  $1\leqslant{}N_{\rm vac}\leqslant{}30$ and two flake sizes $N_{\rm tot}(0)=8278$ and  $N_{\rm tot}(0)=32758$ are displayed in the right panel of Fig.\ \ref{psnvactri}. The dependence of $\lambda_{\rm fit}$ on  $N_{\rm vac}$ and $N_{\rm tot}$ can be rationalize within a~power-law
\begin{equation}
  \label{lamfitnvac}
  \lambda_{\rm fit}\approx{}0.103\times{}
  \left(N_{\rm vac}\sqrt{N_{\rm edge}}\right)^{0.34},
\end{equation}
where the total number of terminal sites 
\begin{equation}
  N_{\rm edge}=3\sqrt{N_{\rm tot}+N_{\rm vac}+3}-3.
\end{equation}

\subsection{\rmfamily Phase diagram for triangular flakes with zigzag edges}
Eq.\ (\ref{lamfitnvac}) is now employed to estimate the maximal system size $N_{\rm tot}$, and the maximal number of edge vacancies $N_{\rm vac}$, for which signatures of TRS breaking still can be identified in the spectrum. This is possible as long as $\lambda_{\rm fit}<\lambda_\star=0.27$ [see Eq.\ (\ref{kaplarg})], as for any $\lambda_{\rm fit}\geqslant{}\lambda_\star$ we have $\overline{\kappa}(\lambda_{\rm fit})=0$ and level-spacing distribution simply evolves from that characterising GOE matrix with approximate twofold degeneracy of each level towards GOE without such a~degeneracy. For instance, we obtain
%\begin{gather}
\begin{align}
  \label{ntotnvac1}
  N_{\rm tot}\lesssim{}9500\ \ \ \  &\text{for}\ \ \ \ N_{\rm vac}=1, \\
  N_{\rm tot}\lesssim{}630\ \ \ \  &\text{for}\ \ \ \ N_{\rm vac}=2, \\
  N_{\rm tot}\lesssim{}130\ \ \ \  &\text{for}\ \ \ \ N_{\rm vac}=3. 
\end{align}
%\end{gather}

On the other hand, system size and the number of energy levels taken into account must be large enough to distinguish between spectral fluctuations of GUE and spectral fluctuations of other ensembles. 

Density of states (per one direction of spin) for bulk graphene reads
\begin{equation}
  \label{rhobulk}
  \rho_{\rm bulk}(E)=\frac{{\cal A}}{\pi{}(\hbar{}v_F)^2}|E|=
  \frac{N_{\rm tot}}{\sqrt{3}\,\pi{}t^2}|E|. 
\end{equation}
The number of energy levels $N_{\rm lev}$ in the interval $(0,E_{\rm max})$ can be approximated by 
\begin{equation}
  \label{nlevem}
  N_{\rm lev}\approx{}\int_0^{E_{\rm max}}\rho_{\rm bulk}(E)dE
  \approx{}0.0919\times{}N_{\rm tot}\left(\frac{E_{\rm max}}{t}\right)^2. 
\end{equation}
Physically, occupying $N_{\rm lev}$ electronic levels above the Dirac point one produces the electric charge $Q=-2eN_{\rm lev}$, resulting in a~typical experimental limit of $E_{\rm max}=0.2-0.3\,$eV for graphene nanostructures on SiO$_2$-based substrates \cite{Hua07}. 

\begin{figure}[!t]
\centerline{\includegraphics[width=0.98\linewidth]{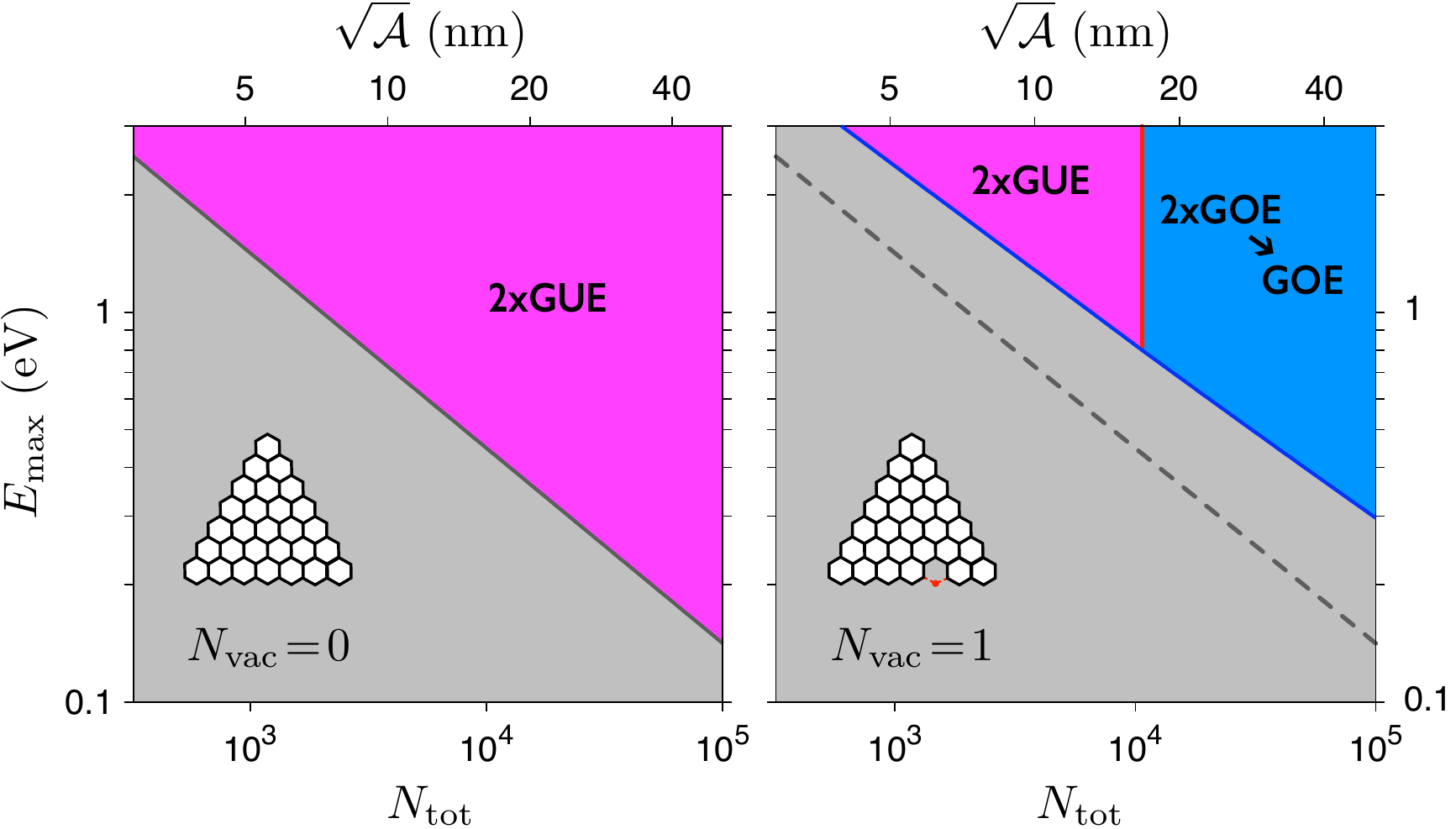}}
\caption{\label{mainfig}
  Phase diagram for triangular graphene nanoflakes with zigzag edges. Grey solid line in left panel (replotted as a~dashed line in right panel) corresponding to Eq.\ (\ref{emntotlam0}) for $N_{\rm vac}=0$ splits the region where number of available energy levels is insufficient to determine the class of spectral fluctuation (below the line) and the region where one should be able to identify the unitary class with approximate twofold degeneracy (2xGUE). Blue solid line in right panel is same as solid line in left panel, but for $N_{\rm vac}=1$, calculated numerically from Eq.\ (\ref{emntot}) for $\lambda=\lambda_{\rm fit}(N_{\rm tot})$ (see Eq.\ (\ref{lamfitnvac})). Vertical red line in right panel marks the limit given by Eq.\ (\ref{ntotnvac1}), above which the orthogonal class with gradual degeneracy splitting (2xGOE$\rightarrow$GOE) appears. 
}
\end{figure}

Level-spacing distributions $P(S)$ are normalized such that $\langle{}1\rangle=\langle{}S\rangle=1$. In turn, the variance $\mbox{Var}\left\{S\right\}\equiv{}\sigma^2=\langle{}S^2\rangle-1$ raises as the lowest moment allowing one to distinguish between different distributions. In particular, we have
\begin{gather}
  \label{valgogu}
  \mbox{Var}\left\{S\right\}_{\rm GOE}=\frac{4}{\pi}-1
  \approx{}0.273, \ \ \ \ \ \ 
  \mbox{Var}\left\{S\right\}_{\rm GUE}=\frac{3\pi}{8}-1
  \approx{}0.178,  \\
  \label{varlam0}
  \text{and}\ \ \ \ \ \ 
  \mbox{Var}\left\{S\right\}_{{\lambda}\rightarrow{}0}=\frac{3\pi}{4}-1
  \approx{}1.356,
\end{gather}
where Eq.\ (\ref{varlam0}) refers to the empirical distribution $P_{\overline{\alpha},\overline{\kappa}}(\lambda,S)$ given by Eq.\ (\ref{psoveralka}) with $\lambda\rightarrow{}0$. Similar calculation for arbitrary $\lambda$ is straightforward, but the resulting formula is too lengthy to be presented.\footnote{We use the property of $m$-th cumulant of the distribution $P(S)=\frac{1}{2}[aP_1(aS)+bP_2(bS)]$, which is equal to $\langle{S^m}\rangle_P=\frac{1}{2}(a^{-m}\langle{}S^m\rangle_{P_1}+b^{-m}\langle{}S^m\rangle_{P_2})$. For $P_1=P_{\rm GOE}$ and $P_2=P_{\rm GOE-GUE}$, see Eqs.\ (\ref{psgoe}) and (\ref{lenzyc}), necessary integrals for $m=2,3,4$ can be calculated analytically.} When  $\mbox{Var}\left\{S\right\}$ is calculates for a~large but finite collection of spacings $N_{\rm spc}=N_{\rm lev}\!-\!1$, it becomes a~random variable itself, with a~variance which can be approximated by
\begin{equation}
  \label{varvars}
  \mbox{Var}\left\{ \mbox{Var}\left\{S\right\} \right\}\approx
  \frac{1}{N_{\rm lev}}\left(\mu_4-\sigma^2\right)=
  \frac{1}{N_{\rm lev}}
  \left(\langle{}S^4\rangle -4\langle{}S^3\rangle +8\langle{}S^2\rangle
  -\langle{}S^2\rangle^2 -4\right),
\end{equation}
where $\mu_4=\langle{}\left(S-\langle{}S\rangle\right)^4\rangle$ denotes the 4-th central moment and we have used the normalization $\langle{}S\rangle=1$. In turn, for spacings following the distribution  $P_{\overline{\alpha},\overline{\kappa}}(\lambda,S)$  (\ref{psoveralka}) one can find they do not follow GOE if 
\begin{equation}
  N_{\rm lev}\gtrsim{}\left[
    \frac{3\sqrt{\left(\mu_4-\sigma^2\right)_{\lambda}}}{\mbox{Var}\left\{S\right\}_{\lambda}-\mbox{Var}\left\{S\right\}_{\rm GOE}}
  \right]^2,
\end{equation}
where the factor $3$ in the nominator corresponds to the $3\sigma$ level of significance. Substituting Eq.\ (\ref{nlevem}) one can rewrite the above as 
\begin{equation}
  \label{emntot}
  E_{\rm max}\sqrt{N_{\rm tot}}\ \gtrsim{}\ 9.9\,t\times
  \frac{\sqrt{\left(\mu_4-\sigma^2\right)_{\lambda}}}{\mbox{Var}\left\{S\right\}_{\lambda}-0.273}.
\end{equation}
For $\lambda\rightarrow{}0$ we have $\mu_4-\sigma^2\rightarrow{}\frac{21}{16}\pi^2-2\pi-4\approx{}2.671$, leading to 
\begin{equation}
  \label{emntotlam0}
%  \left(E_{\rm max}\sqrt{N_{\rm tot}}\right)_{\lambda\rightarrow{}0}\ 
%  \gtrsim{}\ 14.9\,t. 
  N_{\rm tot}\ \gtrsim{}\ 223\times\left(t/E_{\rm max}\right)^2 \qquad 
\text{for}  \quad N_{\rm vac}=0.
\end{equation}

Limiting values of $N_{\rm tot}$ and $E_{\rm max}$, following from Eqs.\ (\ref{ntotnvac1}), (\ref{emntot}), and (\ref{emntotlam0}) are in depicted Fig.\ \ref{mainfig}, presenting the central results of this work. In the absence of edge vacancies ($N_{\rm vac}=0$), the attainable Fermi energy $E_{\rm max}=0.25\,$eV should make it possible to detect TRS breaking in nanoflakes containing $N_{\rm tot}\gtrsim{}3\cdot{}10^4$ carbon atoms, corresponding to the physical diameter of $\sqrt{{\cal A}}\approx{}15\,$nm. For $N_{\rm vac}=1$, the limit of $N_{\rm tot }\lesssim{}9500$ (see Eq.\ (\ref{ntotnvac1})) implies $E_{\rm max}\gtrsim{}0.8\,$eV is required, slightly exceeding current experimental limits for graphene nanostructures.

\section{\rmfamily Concluding remarks}
We have revisited level-spacing statististics of triangular graphene nanoflakes with zigzag edges, subjected to weak substrate-induced disorder, earlier discussed in Ref.\ \cite{Ryc12}. Our previous study is complemented by comparing spectral fluctuations of the systems considered with spectral fluctuations of large random matrices belonging to a~mixed ensemble interpolating between GUE with self-dual symmetry and generic GOE. The results show that for a~fixed value of maximal Fermi energy $E_{\rm max}$ (in typical experiment, the Fermi energy is tuned in the range $-E_{\rm max}<E<E_{\rm max}$ by top gate electrode \cite{Pon08,Pal12}), the system size required to detect signatures of the time-reversal symmetry (TRS) breaking at zero magnetic field is bounded from the bottom by the condition for minimal number of quantum-dot energy levels allowing one to distinguish between different classes of spectral fluctuations. A~finite number of vacancies at the system boundary may lead to intervalley scattering restoring TRS, resulting in additional, upper limit for the system size. 

In conclussion, we expect that triangular graphene flakes with {\em perfect} zigzag edges may show signatures of TRS breaking starting from physical sizes exceeding $15\,$nm. For a~finite number of atomic-scale defects  (starting from a~single edge vacancy), one should  search for signatures of the unitary symmetry class in artificial graphene-like systems \cite{Sin11,Tor12,Gom12} rather then in real graphene nanoflakes.

\section*{\rmfamily Acknowledgments}
I thank to Huang Liang for the correspondence. The work was supported by the National Science Centre of Poland (NCN) via Grant No.\ 2014/14/E/ST3/00256. Computations were partly performed using the PL-Grid infrastructure.

%%%%%%%%%%%%%%%%%%%%%%%%%%%%%%%%%%%%%%%%%%%%%%%%%%%%%%%%%%%%
%%%%%%%%%%%%%%%%%%%%%%%%%%%%%%%%%%%%%%%%%%%%%%%%%%%%%%%%%%%%

\end{document}